\input{aipcheck}

\documentclass[
    ,final            
  ]
  {aipproc}

\layoutstyle{6x9}

\begin{document}

\title{Pulsars with the Australian Square Kilometre Array Pathfinder}


\author{I.~H.~Stairs}{
  address={Dept. of Physics and Astronomy, UBC, 6224 Agricultural
    Road, Vancouver, BC V6T 1Z1 Canada}
}

\author{M.~J.~Keith}{
  address={CSIRO Astronomy and Space Science, Australia Telescope National Facility, CSIRO, PO Box 76, Epping, NSW 1710, Australia}
}

\author{Z. Arzoumanian}{
  address={CRESST/USRA, NASA Goddard Space Flight Center, Code 662, Greenbelt, MD 20771, USA}
}
\author{W.~Becker}{
  address={Max-Planck-Institut für extraterrestrische Physik,
    Giessenbachstrasse 1, 85748 Garching, Germany}
}
\author{A.~Berndsen}{
  address={Dept. of Physics and Astronomy, UBC, 6224 Agricultural
    Road, Vancouver, BC V6T 1Z1 Canada}
}
\author{A.~Bouchard}{
  address={Department of Physics, McGill
    University, 3600 University Street, Montreal, Quebec, H3A 2T8,
    Canada}
}
\author{N.~D.~R.~Bhat}{
  address={Swinburne University of Technology, Centre for Astrophysics
    and Supercomputing, Mail H39, PO Box 218, VIC 3122, Australia}
}
\author{M.~Burgay}{
  address={INAF - Osservatorio Astronomico di Cagliari, localit\`a
    Poggio dei Pini, strada 54, I-09012 Capoterra, Italy}
}
\author{D.~J.~Champion}{
  address={Max Planck Institut f\"ur Radioastronomie, Auf dem H\"ugel 69,
    53121 Bonn, Germany}
}
\author{S.~Chatterjee}{
  address={Astronomy Department, Cornell University, Ithaca, NY 14853,
    USA}
}
\author{T.~Colegate}{
  address={ICRAR/Curtin Institute of Radio Astronomy, GPO Box U1987,
    Perth, WA 6845 Australia}
}
\author{J.~M.~Cordes}{
  address={Astronomy Department, Cornell University, Ithaca, NY 14853,
    USA}
}
\author{F.~M.~Crawford}{
  address={Department of Physics and Astronomy, Franklin and Marshall
    College, P.O. Box 3003, Lancaster, PA 17604, USA}
}
\author{R.~Dodson}{
  address={ICRAR/University of Western Australia, Fairway 7, Crawley,
    Perth, WA 6009 Australia}
}
\author{P.~C.~C.~Freire}{
  address={Max Planck Institut f\"ur Radioastronomie, Auf dem H\"ugel 69,
    53121 Bonn, Germany}
}
\author{G.~B.~Hobbs}{
  address={CSIRO Astronomy and Space Science, Australia Telescope National Facility, CSIRO, PO Box 76, Epping, NSW 1710, Australia}
}
\author{A.~W.~Hotan}{
  address={ICRAR/Curtin Institute of Radio Astronomy, GPO Box U1987,
    Perth, WA 6845 Australia}
}
\author{S.~Johnston}{
  address={CSIRO Astronomy and Space Science, Australia Telescope National Facility, CSIRO, PO Box 76, Epping, NSW 1710, Australia}
}
\author{V.~M.~Kaspi}{
  address={Department of Physics, McGill
    University, 3600 University Street, Montreal, Quebec, H3A 2T8,
    Canada}
}
\author{V.~Kondratiev}{
  address={Netherlands Institute for Radio Astronomy (ASTRON), Postbus
    2, 7990 AA Dwingeloo, The Netherlands}
}
\author{M.~ Kramer}{
  address={Max Planck Institut f\"ur Radioastronomie, Auf dem H\"ugel 69,
    53121 Bonn, Germany}
,altaddress={University of Manchester, Jodrell Bank Centre for
    Astrophysics, Alan Turing Building, Manchester M13 9PL, U.K.}
}
\author{T.~J.~W. Lazio}{
  address={Jet Propulsion Laboratory, M/S 138-308, 4800 Oak Grove Dr.,
    Pasadena, CA 91109, USA}
}
\author{W.~Majid}{
  address={Jet Propulsion Laboratory, M/S 138-308, 4800 Oak Grove Dr.,
    Pasadena, CA 91109, USA}
}
\author{R.~N.~Manchester}{
  address={CSIRO Astronomy and Space Science, Australia Telescope National Facility, CSIRO, PO Box 76, Epping, NSW 1710, Australia}
}
\author{D.~J.~Nice}{
  address={Department of Physics, Lafayette College, Easton, PA 18042,
    USA}
}
\author{A.~Pellizoni}{
  address={INAF - Osservatorio Astronomico di Cagliari, localit\`a
    Poggio dei Pini, strada 54, I-09012 Capoterra, Italy}
}
\author{A.~Possenti}{
  address={INAF - Osservatorio Astronomico di Cagliari, localit\`a
    Poggio dei Pini, strada 54, I-09012 Capoterra, Italy}
}
\author{S.~M.~Ransom}{
  address={National Radio Astronomy Observatory, Charlottesville, VA
    22903, USA}
}
\author{N.~Rea}{
  address={Institut de Ci\`encies de l'Espai (IEEC-CSIC),
    Campus UAB, Facultat de Ci\`encies, Torre C5-parell, 2a planta, 08193
    Barcelona, Spain}
}
\author{R.~Shannon}{
  address={CSIRO Astronomy and Space Science, Australia Telescope National Facility, CSIRO, PO Box 76,
    Epping, NSW 1710, Australia}
}
\author{R.~Smits}{
address={Netherlands Institute for Radio Astronomy (ASTRON), Postbus
   2, 7990 AA Dwingeloo, The Netherlands}
,altaddress={University of Manchester, Jodrell Bank Centre for
    Astrophysics, Alan Turing Building, Manchester M13 9PL, U.K.}
}
\author{B.~W.~Stappers}{
 address={University of Manchester, Jodrell Bank Centre for
    Astrophysics, Alan Turing Building, Manchester M13 9PL, U.K.}
}
\author{D.~F.~Torres}{
  address={ICREA \& Institut de Ci\`encies de l'Espai (IEEC-CSIC),
    Campus UAB, Facultat de Ci\`encies, Torre C5-parell, 2a planta, 08193
    Barcelona, Spain}
}
\author{A.~G.~J.~van~Leeuwen}{
  address={Netherlands Institute for Radio Astronomy (ASTRON), Postbus
    2, 7990 AA Dwingeloo, The Netherlands}
,altaddress={Astronomical Institute ``Anton Pannekoek,'' University of
Amsterdam, 1098 SJ Amsterdam, Netherlands}
}
\author{W.~van~Straten}{
  address={Swinburne University of Technology, Centre for Astrophysics
    and Supercomputing, Mail H39, PO Box 218, VIC 3122, Australia}
}
\author{P.~Weltevrede}{
  address={University of Manchester, Jodrell Bank Centre for
    Astrophysics, Alan Turing Building, Manchester M13 9PL, U.K.}
}

\begin{abstract}
  The Australian Square Kilometre Array Pathfinder (ASKAP) is a
  36-element array with a 30-square-degree field of view being built
  at the proposed SKA site in Western Australia.  We are conducting
  a Design Study for pulsar observations with ASKAP, planning both
  timing and search observations.  We provide an overview of the ASKAP
  telescope and an update on pulsar-related progress.

\end{abstract}

\maketitle


\section{Introduction}

The Australian Square Kilometre Array Pathfinder (ASKAP) is currently
being built at the proposed site for the Square Kilometer Array (SKA)
in Western Australia.  ASKAP will be a 36-element array of 12-m
antennas with baselines of up to several km; it is well-described in
\citet{dgb+09}.  It will use checkerboard-style phased-array feeds \citep{hok+07}
with dithering to achieve roughly uniform sensitivity over a field of
view (FOV) of 30 square degrees.  Several primary beams will be used
to cover the full FOV, with the number of beams increasing with
frequency so as to maintain the same size FOV at all frequencies
between 700\,MHz and 1800\,MHz.  Roughly 300\,MHz of bandwidth will be
available at any time.  ASKAP will also offer multiple tied-array
beams, steerable within the FOV, for projects interested in
point-source work.  The projected system temperature is roughly 50\,K.

The goals of the ASKAP project are to demonstrate technologies
(specifically inexpensive parabolic reflectors and phased-array feeds)
relevant to the SKA, to carry out science related to the SKA Key
Science Projects \citep{cr04}, to establish a radio astronomy site in Western
Australia and to build a user base for the SKA \citep{gjfc08}.

The choice has been made to optimize the combination of antenna
layout, FOV, observing frequencies and system temperature to allow
rapid surveys of neutral Hydrogen in our Galaxy and nearby galaxies, along
with continuum surveys of galaxies to high redshift.  Polarization
surveys and exploration of transient phenomena are also priorities.
The full science case for ASKAP can be found in \citet{jbb+07,jtb+08}.

ASKAP will reduce all imaging data taken with one or more real-time
pipelines, archiving images and spectral lines as required.  Observers
will retrieve data from the archive rather than operating the
telescope themselves.  All archived data will be publically available
\citep{dgb+09}.  To ensure that good analysis pipelines are developed,
observers have been encouraged to organize into ``Science Survey''
teams, 10 of which have been chosen to carry out ``Design Studies'' in
the period 2009-2011.  Advantages of joining a Science Survey Project
(SSP) include ongoing communication with and support from CASS during
ASKAP development, ensuring that SSP scientists will be well-placed to
understand and extract science from the data the moment they appear in
the archive.

With a total collecting area similar to that of the Parkes telescope,
but significantly higher system temperature, ASKAP is perhaps not the
most obvious choice for pulsar observations.  Nevertheless, pulsars
are at the core of one of the 5 SKA Key Science Projects
(``Strong-Field Tests of Gravity Using Pulsars and Black Holes'') and
form one of the two driving projects for the Phase I SKA.  Since
pulsar observations have historically been carried out primarily using
large single dishes, we have formed the COAST (``Compact Objects with
ASKAP: Surveys and Timing'') SSP collaboration to work on the
technical aspects of migrating pulsar observations to interferometers
while carrying out niche pulsar science in the areas where ASKAP can
excel.

\section{The Pulsar Science Case}

The COAST SSP proposal comprises both search and timing observations,
aimed at a broad range of pulsar-related science.

\subsection{Timing}

Despite somewhat lower point-source sensitivity than Parkes (nominally
about 65\%, but it could be about 90\% if the system temperature can
be reduced to 35\,K), ASKAP can still achieve good signal-to-noise
ratios on millisecond pulsars (MSPs) with sufficiently long
integrations. It is in a low-interference environment and should be
somewhat more robust to interference than a single dish. These
properties could allow it to take on part of the Southern-hemisphere
burden of monitoring MSPs for the purposes of contraints on a
background of gravitational waves, and perhaps participate in an
eventual detection \citep[e.g., ][]{jhlm05}.  This would make ASKAP
part of the International Pulsar Timing Array Project.  While the MSPs
are being monitored, the rest of the 30-square-degree ASKAP FOV need
not go to waste; there are often other pulsars visible, and in the
Galactic Plane there are regions where 20 or more pulsars are
accessible at once (though not all of these contain MSPs).  With new
pulsars being discovered in ongoing surveys \citep[e.g., ][]{kjv+10},
the surface density of sources will only increase.  In general, we
expect most of the ``secondary'' pulsars in the FOV to be young.  Many
may glitch or show large amounts of timing systematics \citep[e.g.,
][]{lhk+10} and/or be targets for high-energy observations, for
example with the Fermi Gamma-ray Observatory.  Both young and millisecond pulsars
are exciting targets for gravitational-wave observatories such as
LIGO.  There are therefore multiple motivations for ongoing monitoring
to produce accurate radio ephemerides, and to record time-series data
for pulsars known to glitch.

\subsection{Point-Source Searches}

The myriad science areas accessible to pulsar observations will all
benefit from the discovery of previously unknown pulsars, either by
finding unique objects and test-cases, or by the improved statistics
of a larger and more comprehensive sample.  Again, it is necessary to
iron out the technical details of how to accomplish pulsar surveys
with an interferometer, in order to fulfill part of one of the Key
Science Projects of the SKA \citep{kbc+04}.

One technique aimed at discovering primarily millisecond and fast
binary pulsars will be to search pulsar-like point sources found in
the large continuum surveys such as EMU (Evolving Map of the Universe)
and the transient surveys such as VAST (Variability and Slow
Transients) and CRAFT (Commensal Real-time ASKAP Fast Transients)
that will be carried out with ASKAP.  EMU will find weak point
sources, which we can examine for steep spectral indices and/or strong
polarization, which may be signposts to pulsars.  VAST and CRAFT will
find stronger sources that are time- and/or frequency-variable; the
scintillation properties may also indicate that pulsars are present.
Work is underway to examine current point-source catalogs to determine
the optimal criteria to use.  Using tied-array beams on the point
sources, we will be able to record fast-sampled,
high-frequency-resolution data and thus be sensitive to millisecond
pulsars.  The selection effects will be different from those of
traditional wide-area searches, possibly opening up an entirely new
pulsar discovery space.

\subsection{Wide-Area Searches}

Once again aiming to develop techniques that will be needed for the
SKA, we also plan to carry out a search of a wide area of sky, using
the data recorded from the entire ASKAP FOV.  Given the data rate
limitations from the correlator, such a search will necessarily be
restricted to comparatively wide frequency channels of a few MHz and
to sample rates of milliseconds; even then the development of the
necessary correlator modes will be some years off.  The data reduction
challenge is also significant, given the need to search either every
pixel in the FOV or else the UV plane itself.  We may choose to use
only the inner core of 30 antennas to restrict the baselines to 2\,km
and reduce the number of pixels.

This search will of necessity be sensitive primarily to slower pulsars
with spin periods greater than $\sim 10$ milliseconds, but there are a
number of exciting types of objects in this category, including young
pulsars, radio magnetars and even double-neutron-star binaries.  To
maximize the scientific output from such a survey, we intend to
conduct searches in the lowest part of the ASKAP frequency band
(700--1000\,MHz, a regime not considered so far in wide-area searches)
and a target zone just off the Galactic plane to minimize dispersion
and scattering, while expecting both mildly recycled pulsars and young
pulsars to be in our fields.  Another likely target for such a survey
will be the Magellanic Clouds, which we can cover deeply in just one
pointing each.

\section{Technical Developments and Plans}

Through the Design Study, we are investigating possible solutions to
several algorithmic and hardware problems as well as planning trial
observations with the BETA (Boolardy Engineering Test Array) prototype
of 6 antennas that should be available in 2011-2. 

Although the timing science plan is well-understood, one area that is
being investigated is the optimal selection of particular fields on
the sky that will place the most interesting pulsars near the centre
of the FOV to minimize polarization impurities while still maintaining
a good number of interesting secondary pulsars.  Since we expect the list
of primary pulsars to evolve as more objects are discovered, we need a
reliable algorithm for FOV selection rather than a rigid list.

\begin{figure}[ht]
 \includegraphics[height=.4\textheight,type=eps,ext=.eps,read=.eps]{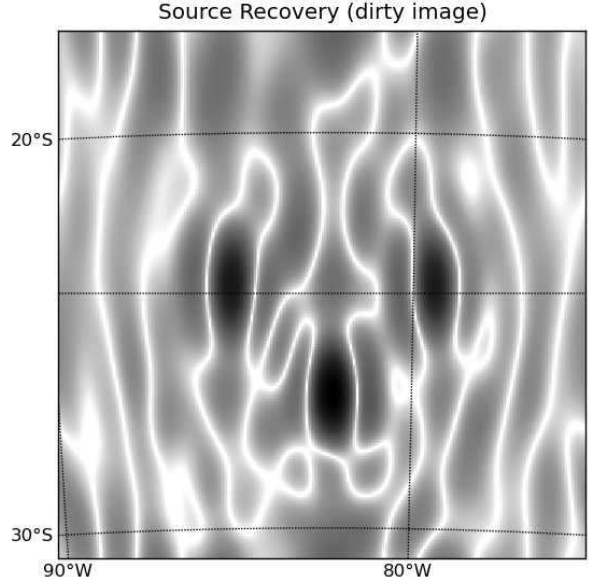}
 \caption{``Dirty image'' showing recovery of simulated point sources. Visibility
data was generated from an input sky with three equal-strength point
sources, as observed by BETA tracking the southernmost source. The
system temperature was set to 10 percent of the source flux.} \label{fig:source}
\end{figure}

 The point-source-search selection criteria are under active
exploration using existing catalogs from the ATLAS survey
\citep{naa+06}.  Population syntheses for both the point-source and
wide-area surveys are being carried out with two different code
packages; see the article by Bouchard in this volume for details.
Meanwhile, algorithm development for the wide-area searches is
proceeding.  Currently this includes simulation of both point-like and
pulsed sources on the sky and software to detect pulsations along
individual baselines.  The next steps include searches of the
simulated pixelized sky and development of an algorithm to indentify
pulsar signals from unknown sky positions using multiple baselines
simultaneously, effectively searching directly in the UV plane.  The
success of the intermediate stages of this software are shown in
Figure~\ref{fig:source}, which demonstrates the successful recovery of
simulated point sources.

We will need to provide pulsar backend hardware to acquire the signals
and process them.  This is made significantly easier relative to what
is required at most telescopes by the ASKAP engineers' intention to
provide us with tied-array beam signals that are already
Nyquist-sampled and separated into few-MHz channels, or else in a
full-Stokes digitized filterbank format.  Therefore most of our data
acquisition needs can be satisfied by computing power to handle the
coherent dedispersion task for MSP timing, to fold or make dedispersed
timeseries data for the slower pulsars and simply to record the
filterbank data for the point-source searches.  Disk space will be the
main requirement for the last type of data, while small (possibly
GPU-based) clusters should be able to handle the computational
tasks. The wide-area survey will require significantly more and faster
disks; this problem will be addressed closer to the time when these
observations might be scheduled.

Our observations with the 6-element BETA telescope will focus on
timing several bright pulsars simultaneously.  In particular, we
intend to observe two pulsars separated across the
field of view to characterize the stability of the telescope response
and tied-array beams in particular over data spans of several hours.




\begin{theacknowledgments}
The Murchison Radio-astronomy Observatory (MRO) is jointly funded by the
Commonwealth Government of Australia and State Government of Western
Australia and managed by the CSIRO. We acknowledge the Wajarri
Yamatji people as the traditional owners of the Observatory site.
 Pulsar and ASKAP research at UBC is supported by NSERC
Discovery and SRO grants.
\end{theacknowledgments}





\begin{thebibliography}{11}
\expandafter\ifx\csname natexlab\endcsname\relax\def\natexlab#1{#1}\fi
\providecommand{\enquote}[1]{``#1''}
\expandafter\ifx\csname url\endcsname\relax
  \def\url#1{\texttt{#1}}\fi
\expandafter\ifx\csname urlprefix\endcsname\relax\def\urlprefix{URL }\fi
\providecommand{\eprint}[2][]{\url{#2}}

\bibitem[{Deboer} et~al.(2009)]{dgb+09}
D.~R. {Deboer}, R.~G. {Gough}, J.~D. {Bunton}, T.~J. {Cornwell}, R.~J.
  {Beresford}, S.~{Johnston}, I.~J. {Feain}, A.~E. {Schinckel}, C.~A.
  {Jackson}, M.~J. {Kesteven}, A.~{Chippendale}, G.~A. {Hampson}, J.~D.
  {O'Sullivan}, S.~G. {Hay}, C.~E. {Jacka}, T.~W. {Sweetnam}, M.~C. {Storey},
  L.~{Ball}, and B.~J. {Boyle}, \emph{IEEE Proceedings} \textbf{97}, 1507--1521
  (2009).

\bibitem[Hay et~al.(2007)]{hok+07}
S.~G. Hay, J.~D. O'Sullivan, J.~S. Lot, C.~Granet, A.~Grancea, A.~R. Forsyth,
  and D.~H. Hayman, \enquote{{Focal Plane Array Development for ASKAP
  (Australian SKA Pathfinder)},} in \emph{Proc. EUCAP'07}, 2007.

\bibitem[{Carilli} and {Rawlings}(2004)]{cr04}
C.~{Carilli}, and S.~{Rawlings}, \emph{astro-ph/0409274}  (2004).

\bibitem[Gupta et~al.(2008)]{gjfc08}
N.~Gupta, S.~Johnston, I.~Feain, and T.~Cornwell  (2008), {CSIRO} internal
  document, available at http://www.atnf.csiro.au/SKA/newdocs/configs-3.pdf.

\bibitem[{Johnston} et~al.(2007)]{jbb+07}
S.~{Johnston}, M.~{Bailes}, N.~{Bartel}, C.~{Baugh}, M.~{Bietenholz},
  C.~{Blake}, R.~{Braun}, J.~{Brown}, S.~{Chatterjee}, J.~{Darling},
  A.~{Deller}, R.~{Dodson}, P.~G. {Edwards}, R.~{Ekers}, S.~{Ellingsen},
  I.~{Feain}, B.~M. {Gaensler}, M.~{Haverkorn}, G.~{Hobbs}, A.~{Hopkins},
  C.~{Jackson}, C.~{James}, G.~{Joncas}, V.~{Kaspi}, V.~{Kilborn},
  B.~{Koribalski}, R.~{Kothes}, T.~L. {Landecker}, E.~{Lenc}, J.~{Lovell},
  J.~{Macquart}, R.~{Manchester}, D.~{Matthews}, N.~M. {McClure-Griffiths},
  R.~{Norris}, U.~{Pen}, C.~{Phillips}, C.~{Power}, R.~{Protheroe},
  E.~{Sadler}, B.~{Schmidt}, I.~{Stairs}, L.~{Staveley-Smith}, J.~{Stil},
  R.~{Taylor}, S.~{Tingay}, A.~{Tzioumis}, M.~{Walker}, J.~{Wall}, and
  M.~{Wolleben}, \emph{Proc. Astr. Soc. Aust.} \textbf{24}, 174--188 (2007).

\bibitem[{Johnston} et~al.(2008)]{jtb+08}
S.~{Johnston}, R.~{Taylor}, M.~{Bailes}, N.~{Bartel}, C.~{Baugh},
  M.~{Bietenholz}, C.~{Blake}, R.~{Braun}, J.~{Brown}, S.~{Chatterjee},
  J.~{Darling}, A.~{Deller}, R.~{Dodson}, P.~{Edwards}, R.~{Ekers},
  S.~{Ellingsen}, I.~{Feain}, B.~{Gaensler}, M.~{Haverkorn}, G.~{Hobbs},
  A.~{Hopkins}, C.~{Jackson}, C.~{James}, G.~{Joncas}, V.~{Kaspi},
  V.~{Kilborn}, B.~{Koribalski}, R.~{Kothes}, T.~{Landecker}, A.~{Lenc},
  J.~{Lovell}, J.~{Macquart}, R.~{Manchester}, D.~{Matthews},
  N.~{McClure-Griffiths}, R.~{Norris}, U.~{Pen}, C.~{Phillips}, C.~{Power},
  R.~{Protheroe}, E.~{Sadler}, B.~{Schmidt}, I.~{Stairs}, L.~{Staveley-Smith},
  J.~{Stil}, S.~{Tingay}, A.~{Tzioumis}, M.~{Walker}, J.~{Wall}, and
  M.~{Wolleben}, \emph{Experimental Astronomy} \textbf{22}, 151--273 (2008).

\bibitem[{Jenet} et~al.(2005)]{jhlm05}
F.~A. {Jenet}, G.~B. {Hobbs}, K.~J. {Lee}, and R.~N. {Manchester}, \emph{ApJ}
  \textbf{625}, L123--L126 (2005).

\bibitem[{Keith} et~al.(2010)]{kjv+10}
M.~J. {Keith}, A.~{Jameson}, W.~{van Straten}, M.~{Bailes}, S.~{Johnston},
  M.~{Kramer}, A.~{Possenti}, S.~D. {Bates}, N.~D.~R. {Bhat}, M.~{Burgay},
  S.~{Burke-Spolaor}, N.~{D'Amico}, L.~{Levin}, P.~L. {McMahon}, S.~{Milia},
  and B.~W. {Stappers}, \emph{MNRAS} \textbf{409}, 619--627 (2010).

\bibitem[{Lyne} et~al.(2010)]{lhk+10}
A.~{Lyne}, G.~{Hobbs}, M.~{Kramer}, I.~{Stairs}, and B.~{Stappers},
  \emph{Science} \textbf{329}, 408--412 (2010).

\bibitem[Kramer et~al.(2004)]{kbc+04}
M.~Kramer, D.~C. Backer, J.~M. Cordes, T.~J.~W. Lazio, B.~W. Stappers, and
  S.~Johnston, \emph{New Astr.} \textbf{48}, 993--1002 (2004).

\bibitem[{Norris} et~al.(2006)]{naa+06}
R.~P. {Norris}, J.~{Afonso}, P.~N. {Appleton}, B.~J. {Boyle}, P.~{Ciliegi},
  S.~M. {Croom}, M.~T. {Huynh}, C.~A. {Jackson}, A.~M. {Koekemoer}, C.~J.
  {Lonsdale}, E.~{Middelberg}, B.~{Mobasher}, S.~J. {Oliver}, M.~{Polletta},
  B.~D. {Siana}, I.~{Smail}, and M.~A. {Voronkov}, \emph{AJ} \textbf{132},
  2409--2423 (2006).

\end{thebibliography}

\IfFileExists{\jobname.bbl}{}
 {\typeout{}
  \typeout{******************************************}
  \typeout{** Please run "bibtex \jobname" to optain}
  \typeout{** the bibliography and then re-run LaTeX}
  \typeout{** twice to fix the references!}
  \typeout{******************************************}
  \typeout{}
 }

\end{document}